\documentclass[sigconf]{acmart}

\AtBeginDocument{%
  \providecommand\BibTeX{{%
    \normalfont B\kern-0.5em{\scshape i\kern-0.25em b}\kern-0.8em\TeX}}}

\setcopyright{acmcopyright}
\copyrightyear{2022}
\acmYear{2022}
\acmDOI{10.1145/1122445.1122456}
\acmConference[VU '22]{Vrije Universiteit Amsterdam}{14-16 June, 2022}{Amsterdam, The Netherlands}
\acmBooktitle{Vrije Universiteit Amsterdam, 2022, Amsterdam, The Netherlands}
\acmPrice{15.00}
\acmISBN{978-1-4503-XXXX-X/18/06}

\usepackage{pifont}
\usepackage{colortbl,multirow,hhline}
\usepackage{float}
\usepackage{tablefootnote}
\usepackage{enumitem}
\usepackage{longtable}
\usepackage[most]{tcolorbox}
\usepackage{lipsum}
\usepackage{tablefootnote}
\usepackage{pifont}

\title{Architecture Knowledge Representation and Communication Industry Survey}
\author{Haben Birhane Gebreweld}
\affiliation{%
   \institution{Vrije Universiteit Amsterdam}
   \country{The Netherlands}}
\email{h.gebreweld@student.vu.nl}

\begin{document}

\begin{abstract}
 \textbf{\textit{Background: }} The literature presents several approaches, such as views, viewpoint, and architecture decision records (ADRs), to describe software architectural knowledge (AK). On the other hand, sustainability is a subject that is receiving increasing attention in software engineering, particularly in relation to software architecture. However, there appears to be a lack of industry reviews on these topics from a practical perspective.
 
 \textbf{\textit{Aim: }} 
In this research we aim to understand the current practice in architecture knowledge, and to explore where sustainability can be applied to address sustainability in software architecture in the future.

\textbf{\textit{Method: }}We used a survey, which utilized a questionnaire containing 34 questions and collected responses from 45 architects working at a prominent bank in the Netherlands, aimed to evaluate the practical representation and communication of architectural knowledge and sustainability.

\textbf{\textit{Result: }}Our analysis yielded two primary discoveries and several intriguing detailed results regarding how AK is captured and conveyed to diverse stakeholders. The report aims to communicate two essential messages to guide future research in the field. Firstly, it seems crucial to develop a new architectural element that connects various architectural features and perspectives tailored for different stakeholders. Secondly, providing clear guidance, references, and goals is essential to motivate architects to adopt Sustainable Software Engineering practices.

\textbf{\textit{Conclusion: }}After analysing the data collected through this survey, we have concluded that: \textbf{a)} There are no established domain-specific AK methods/tools in the financial domain. Most practitioners use domain-generic tools. 
\textbf{b)} A new architectural element that links the various architectural features and viewpoints created for various stakeholders appears to be necessary. 
\textbf{c)} There is sufficient sustainability awareness and motivation among software architects. However, what they lack are clear guidance, references, and goals to practice sustainable software engineering.

\end{abstract}



\maketitle

\section{Introduction}
\label{sec:intro}
Software architectural knowledge refers to the knowledge acquired while designing a software architecture, encompassing the assumptions, decisions, context, and other factors involved in that process \cite{babar2009software}. Various approaches have been developed both in literature and industry to depict this knowledge, such as views and viewpoints\cite{clements2003documenting}, architecture decision records (ADRs)\cite{smith2020getting}, and standards like ISO 42010\cite{clements2003documenting} and the C4 Model\cite{brown2018c4}.
However, for this knowledge to be effective, it is important that all relevant stakeholders share information about the architecture and how it is represented. The way this information is communicated depends on the organizational structures involved and can take various forms, such as wikis, workshops, emails, etc. Understanding how architectural knowledge is represented and communicated in professional practice is important to identify appropriate relationships that address sustainability elements in software architecture. By studying how this knowledge is represented and shared, we can gain insights into best practices for ensuring that this knowledge is effectively communicated and can be used to make informed decisions about the sustainability of software architecture.

As researchers, we develop intriguing methods and techniques for managing architectural knowledge, while practitioners have their preferred means of capturing and sharing architectural information. To ensure that our methods do not become a "silver bullet" that practitioners do not utilize, it is crucial to conduct industry reviews. By building upon existing industry practices and filling in any missing pieces, we can develop effective and useful methods that practitioners will embrace.

The purpose of this research is to gain insight into the current practices related to architecture knowledge and explore how sustainability can be integrated into software architecture in the future. Our objective is to characterize architecture knowledge and sustainability from the perspective of software architects, specifically with regards to representation and communication in the professional practice context. To achieve this, we conducted a questionnaire survey and gathered responses from architects working at a prominent bank in the Netherlands about their experiences in the industry, focusing on how they represent and communicate architectural knowledge and sustainability. Regarding scientific contributions, as far as we are aware, our study is the first of its kind to explore how software architects perceive and utilize sustainability aspects within software architecture in an industrial context, along with examining architectural knowledge representation and communication.

This study offers several significant contributions, including:
\begin{itemize}
    \item[\ding{51}] A practical review of architectural knowledge representation and communication techniques utilized in the industry.
    \item[\ding{51}] An assessment of how practitioners approach representing and communicating sustainability aspects in software architecture.
    \item[\ding{51}] This study presents a particular collection of AK representation and communication techniques utilized by software architects who work in the financial industry.
\end{itemize}    

The paper is structured in the following manner. In section \ref{sec:related}, we review previous studies that have explored the relationship between architectural knowledge and sustainability in software architecture. Section \ref{sec:design} outlines how we designed and executed the survey. We present a summary of the survey results in section \ref{sec:results}, and in section \ref{sec:discussion}, we provide a detailed analysis of the findings. This analysis aims to make sense of the results and convey the main insights gained from the study. Finally, in section \ref{sec:validity_threat}, we discus the threats to validity of our study, and we provide our conclusion in section \ref{sec:conclusion}. 

\section{Related Work}
\label{sec:related}

There is a wide range of architectural modeling languages available, but it is unclear whether they are capable of adequately describing software architecture to meet users' requirements. Additionally, the strengths, limitations, and needs of these languages are uncertain, creating a gap between what is offered and what users need. Malavolta et al.\cite{malavolta2013industry} aimed to address this gap by surveying 48 practitioners from 40 IT businesses in 15 countries in order to plan for the next generation of these languages. The study examined the perceived benefits, drawbacks, and requirements related to existing languages. While practitioners were generally satisfied with the design capabilities of their employed architectural languages, the study revealed dissatisfaction with the architectural language analysis features and their capacity to describe extra-functional attributes. Moreover, the study found that the use of architectural languages in practice was mostly influenced by industry development rather than academic research, and there was a need for a more formal and practical architectural language. Our research shares similarities with the aforementioned study as we, too, are investigating how architectural knowledge is represented from the perspective of industry practitioners. To achieve this, we conducted a survey among software architects working for a leading bank in the Netherlands. Our study is distinct in that it delves into sustainability and the communication of architectural knowledge among stakeholders. In addition to exploring a specific domain (financial domain), we go beyond the use of architecture description languages and investigate how architects communicate and share their knowledge. 
\newline

Despite a significant amount of research and development of various models and tools, the widespread adoption of Architectural Knowledge Management (AKM) in software companies is lacking due to the cost of capturing AK\cite{capilla2008effort} Capilla et al.,. Determining what the industry needs from AK to get through this barrier and identifying the advantages and disadvantages of the current AK techniques are therefore necessary. Capilla et al.\cite{CAPILLA2016191} undertook an informal retrospective analysis based on their prior work as researchers and proponents of numerous AK research methodologies in order to address this. By conducting a series of interviews with various software businesses, they also looked into the trends and problems for a future research agenda to support the usage of AK in contemporary software development methods. They came up with some interesting observations using this method. While we are also looking into the tools and techniques practitioners use to capture and communicate architectural knowledge, which will help us understand current trends in the industry, our study has some parallels to the research mentioned above. Our research, in contrast to the aforementioned study, has a keen focus on comprehending how software architects represent and communicate both architectural knowledge and sustainability. 
\newline

While there have been secondary research studies conducted on sustainability in software engineering, none have particularly addressed software architecture. Andrikopoulos et al.'s\cite{andrikopoulos2022sustainability} systematic mapping study seeks to fill this research gap by exploring the confluence between sustainability and software architecture. The study's findings showed that current studies have neglected the holistic viewpoint required to resolve such a complex issue by excessively emphasizing on particular sustainability-related dimensions. To develop the maturity of the field, more reflective research studies and improved coverage of the architecting life cycle activities are required. The report suggests a research agenda for sustainability-aware software architecture based on these findings. Our research is similar to the study as we also aim to explore the incorporation of sustainability aspects into software architecture. However, our study takes a unique approach by focusing on how sustainability aspects of software can be effectively represented and communicated from the perspective of software architecture practitioners by conducting industry survey with software architects. 

\section{Study design and Execution}
\label{sec:design}
By conducting this study, we aim to provide software architects and the research community with a useful evaluation of how Architecture Knowledge (AK) and Sustainability are represented and communicated from a practical point of view. Our research objective is formally characterized by employing the structure proposed by Basili et al. \cite{gqm} as follows.

\begin{center}
\begin{tabular}{ l l }
 \textit{Analyze} & Architecture Knowledge \& Sustainability \\ 
 \textit{For the purpose of} & Characterizing \\  
 \textit{With respect to} & Representation \& Communication \\
 \textit{From point of view of} & Software Architects \\
 \textit{In the context of} & Professional Practice
\end{tabular}
\end{center}

Therefore, the present study defines the primary research questions (RQs) and its corresponding sub-research questions as follows:
\newline

\begin{enumerate}[leftmargin=*,labelindent=1em]
    \item[$RQ1$] \textit{How is software architecture knowledge represented and communicated in practice?}\\

    When addressing \textit{RQ1}, we investigate and evaluate the entire industrial context of how AK is represented and communicated. In section \ref{sec:results}, we will provide additional information on the tools, techniques, standards, and documentation utilized by software architects in the industry. This will aid us in understanding the industrial structure of AK representation and communication, which we can exploit to integrate sustainability elements into software architecture in the future.
    \newline
    \begin{enumerate}[leftmargin=*,labelindent=1em]
        \item[$RQ1.1$] \textit{How is software architecture knowledge represented in the financial domain?}\\

        Our goal through \textit{RQ1.1} is to achieve a more comprehensive understanding of the frequently utilized and advantageous AK elements in the financial domain. Moreover, we intend to gain insight into the tools utilized in the financial industry to supplement the AK representation component of \textit{RQ1}.
        \newline
        \item[$RQ1.2$] \textit{How is software architecture knowledge communicated in the financial domain?}\\

        Similar to \textit{RQ1.1}, our objective with \textit{RQ1.2} is to gain an understanding of the tools and techniques utilized by practitioners in the financial industry to communicate AK effectively.
        \newline
        \item[$RQ1.3$] \textit{What architecture knowledge methods are domain-specific or domain-generic?}\\

        Our aim with \textit{RQ1.3} is twofold: to identify any architecture knowledge methods that are specific to the financial sector and to identify any domain-generic methods that can be applied to the financial industry.
        \newline
    \end{enumerate}
    
    \item[$RQ2$] \textit{How can sustainability aspects be represented and communicated in software architecture?}\\
    \newline
    Through \textit{RQ2}, we aim to gather insights from software architects on how they would incorporate sustainability aspects into their daily work and software architecture. Given their wealth of expertise, we aim to challenge software architects to provide possible ways for integrating sustainability into software architecture.
\end{enumerate}

\begin{figure}[htbp]
\center
\includegraphics[width=0.85\columnwidth]{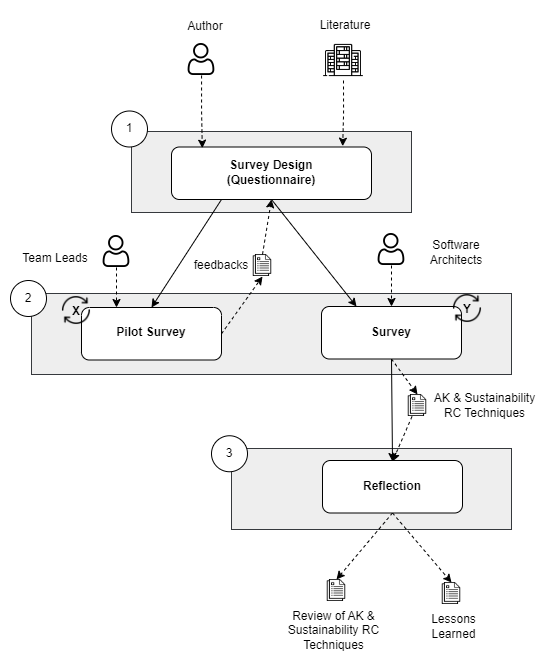}
\caption{Study Design.}
\label{fig:study-design}
\end{figure}

We provide a three-phase research methodology in order to respond to these RQs. Figure \ref{fig:study-design} shows an overview, which is expanded upon below.

In \textbf{Step (1)}, To begin designing the survey, the first step is to identify the research questions and map them to a set of questions that will be used to gather information from software architects about their industry practices. To accomplish this, we have included a variety of questions, including commonly used ones from the literature that pertain to demographics and architectural knowledge. This initial step is crucial because it impacts the quality of the information we obtain from the population. Upon completion, we will have a comprehensive survey that will be distributed to the significant community of software architects at one of the largest banks in the Netherlands. 

We ask respondents \textbf{34 questions} in this survey, with 91\% of them being open-ended and allowing respondents to candidly express their unique experiences. Figure \ref{fig:Suvey_Flow} depicts the flow of the eight blocks numbered Q1-Q8. The blocks in Figure \ref{fig:Suvey_Flow} are labeled with the research questions they intend to answer, except for the consent, demographics, and conclusion blocks. The following blocks are included:

\begin{figure*}[hbt!]
  \includegraphics[scale=0.6]{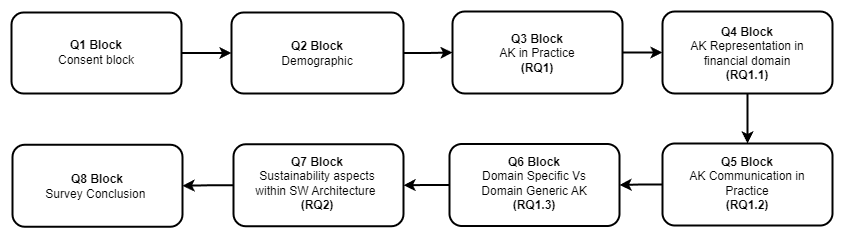}
  \caption{Survey Questionnaire Flow}
  \label{fig:Suvey_Flow} 
\end{figure*}

\textbf{Consent:} We explained the purpose of the study, which is to understand the current practice in architecture knowledge, and to explore where sustainability can be applied to address sustainability in software architecture in the future. We include the researchers' contact details as well as their confidentiality policy. 

\begin{figure*}[hbt!]
  \includegraphics[scale=0.6]{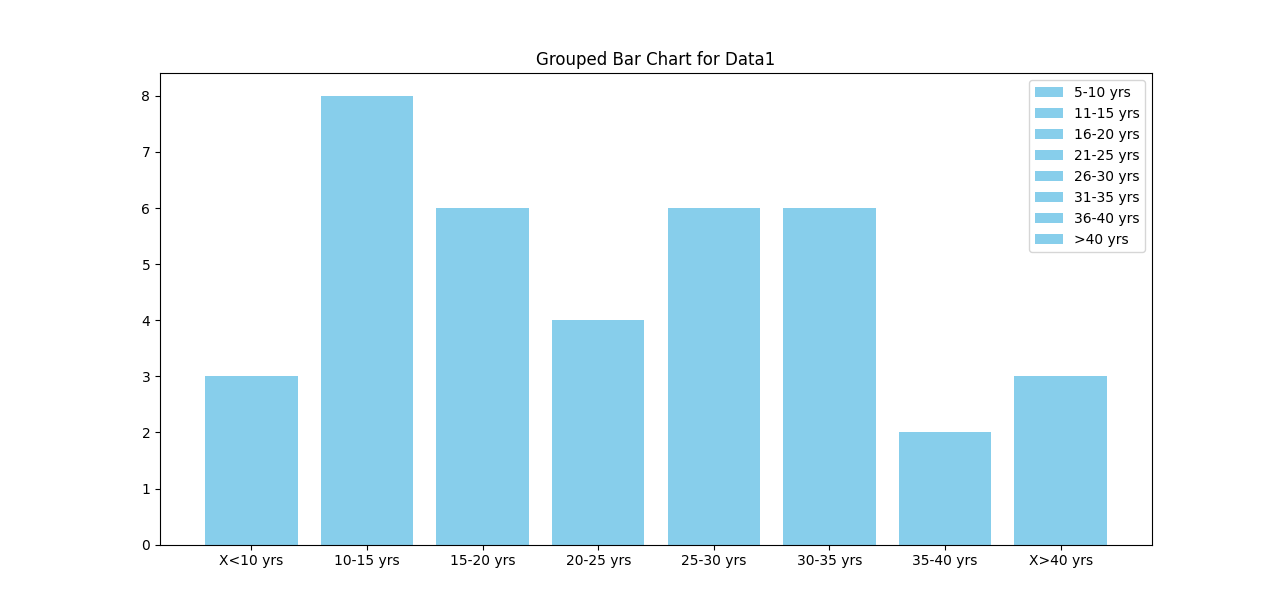}
  \caption{Survey Population Experience in Years }
  \label{fig:Population_Exp} 
\end{figure*}

\begin{figure}[htbp]
\center
\includegraphics[width=0.9\columnwidth]{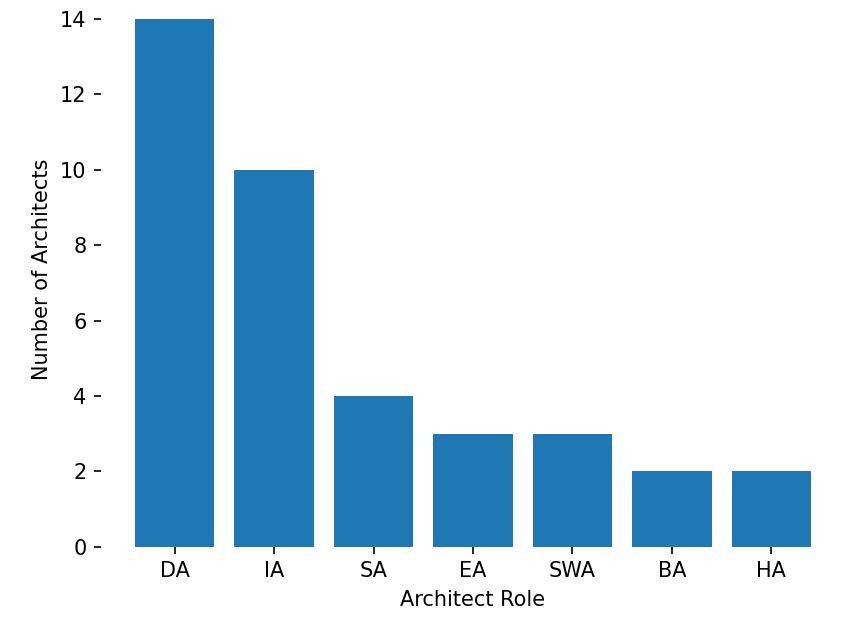}
\caption{ Current Roles of Participants within the bank.\newline \textit{ Where \textbf{DA} stands for domain architect, \textbf{IA} for IT architect, \textbf{SA} for solution architect, \textbf{EA} for enterprise architect, \textbf{SWA} for software architect, \textbf{BA} for business architect,    and \textbf{HA} for hybrid cloud architect.}}
\label{fig:role-figure}
\end{figure}

\textbf{Demographics (Q1.1-Q1.4):} By probing the participants' professional backgrounds, particularly their involvement in software projects (see Figure \ref{fig:Population_Exp}), their experiences in their present organizations, and their specific roles or positions within the organization, this section aims to learn more about the participants (see Figure \ref{fig:role-figure}) 

\textbf{AK in Practice (Q2.1-Q2.5):} We start this part with a formal explanation of our interpretation of architectural knowledge to avoid any misunderstandings (AK)\footnote{\textit{Architecture Knowledge (AK)} is defined as knowledge elements including architecture design, design decisions, assumptions, context, and other elements that together determine why a particular solution is the way it is}. We acknowledge that there may be different ways to understand AK, though. In this section, we are therefore asking participants about the kinds of AK they document and retain, the AK components they believe are missing, and the AK elements they consider to be the most valuable. Our objective in this section is to comprehend how participants are representing and communicating AK. 

\textbf{AK representation in financial domain (Q3.1-Q3.4):} In a similar manner, we began this section by providing a formal description of our interpretation of AK representation\footnote{\textit{Architecture Knowledge (AK) Representation} is defined as capturing and preserving AK in a particular form (e.g., diagrams, PowerPoint, views, viewpoints, or principles)}. We ask participants about the notations, languages, methods, and tools they use to capture AK in general and specifically in their most recent project, as well as the architectural documentation they find most useful. As we are specifically targeting architects who work in a bank, our goal is to gain an understanding of how AK is represented in the financial domain.

\textbf{AK communication in Practice (Q4.1-Q4.3):} In this section, similar to the one above, we provide a description of what is meant by AK communication\footnote{\textit{Architecture Knowledge (AK) Communication} describes to how the knowledge is disclosed between involved stakeholders (e.g., via workshops, or corporate sharing platforms)}. We also inquire with the participants about the stakeholders involved in their roles, as well as the tools and methods they use to communicate with different stakeholders. By doing so, we are able to gather first-hand information on how AK is communicated in practice.

\textbf{Domain Specific Vs Domain Generic AK (Q5.2-Q5.3):} We inquired with the participants about their familiarity with certain AK methods that are unique to their specific business domain, as well as the regulations they keep in mind while representing or communicating AK within their domain. Our goal is to distinguish between AK tools and methods that are specific to their business domain and those that have a more general purpose. 

\textbf{Sustainability aspects within software architecture (Q6.1-Q6.5):} In this section of the survey, our goal is to explore how software practitioners incorporate sustainability aspects into their architectural decisions. To achieve this, we have included a series of questions designed to better understand the participant's perception of IT sustainability. We begin by asking what the concept of IT sustainability means to the participant. Following this, we ask architects whether they consider sustainability aspects during their work, and depending on their response, we delve further to understand which aspects they consider or the reasons for not considering them. We also ask whether participants are aware of any IT sustainability targets set by their company or department, and if they integrate sustainability aspects into their daily work. Through these questions, we seek to gain insights into how architects interpret sustainability, both generally and specifically in the context of software architecture.

\textbf{Survey Conclusion (Q7.1-Q7.2):} Participants are encouraged to provide any additional information they feel is important in this section. Specifically, we inquire about what they believe is necessary to accurately represent and communicate AK, and whether they have any comments about the study itself. These questions are designed to capture any issues that may be of significance to participants but have not been addressed in the survey.  

In \textbf{Step (2)}, we first conduct a pilot survey with a small group of research and industry experts to check the quality of the survey and eliminate any possible pitfalls and unclear aspects before reaching out to the main survey population. This step generates a set of feedback that needs to be incorporated into the original survey design. Then, we conduct a survey with the main population consisting of software architects working for a leading bank in the Netherlands. The objective is to produce a set of architecture and sustainability representation and communication (ARC) techniques used in industry. To \textbf{determine the main population} of the survey and ensure our survey was conducted effectively, we first established the objective of the study, which was to conduct a practical review of architectural knowledge and how it is represented and communicated in the financial industry. After analyzing the possible population for our survey, we determined that software architects would be the most suitable participants. This is because they possess extensive knowledge about AK and its application in the industry (check Figure \ref{fig:Population_Exp}), and because the organization already has architects in different software architecture roles with decades of experience (check Figure \ref{fig:role-figure}).

Next, we reached out to the identified population using a two-fold approach. The first approach involved obtaining a mailing list of software architects within the organization, while the second involved requesting Team leads of different structures within the organization to provide us with a list of architects under their department. By consolidating the information from these two sources, we generated a draft list  of 145 architects. After eliminating redundant names and removing the architects needed for an extensive interview, we arrived at a set of 124 architects. Using these two steps to the best of our abilities, we tried to reach out to all software architects working in the bank. However, we did not conduct any further steps to verify if we indeed reached out to all architects.  

The survey was conducted using \textbf{Qualtrics}\footnote{https://www.qualtrics.com/} and designed to be anonymous to alleviate concerns about judgment or consequences. We reached out to the 124 architects via email and received 45 (39\%) survey responses, with eight indicating they no longer worked at the company and 69 not replying. 

We compiled all recorded responses from all participants into a spreadsheet format where each column represents a specific question and each row represents a response from a participant. However, we made the decision to make all questions, except for the demographic and consent questions, optional. As the consent questions are crucial to obtain legal permission from participants to record their responses, and the demographic data is essential for interpreting the remaining optional questions. As a result, there are variations in the responses we received from each participant, as some attempted to answer all questions, while others chose to skip questions they did not want to answer. 

We analyzed the responses given to each question to find trends and prevailing viewpoints about the specific topic raised by the question. To facilitate analysis, we divided the questions into two categories: closed-ended and open-ended. For closed-ended questions, we simply counted the number of occurrences. For open-ended questions, we categorized responses by interpreting each response's intent(we presented the complete analysis of the responses in the replication package\footnote{\url{https://docs.google.com/spreadsheets/d/1KIMpIWXGCASJX2WlJROAGwDfn1b9yH_p/edit?usp=share_link&ouid=117224618047071995271&rtpof=true&sd=true}}). We created these categories using information from various sources, including the literature (such as the dimensions of sustainability), the study question, and similarities in the responses' intended purposes(check Table \ref{tab:description_of_coding_categories}). Our main objective was to encode open-ended responses so that quantitative and conventional statistical analyses \cite{miles1994} could be applied to the data. 

The reflection of the results to address our core research questions is the final step in \textbf{Step (3)}. Our objective is to provide a summary of the methods for architectural representation and communication that are currently used, as well as how architects address sustainability issues when designing software. We end by summarizing the takeaways from this practical review. 

Based on the data presented in Figure \ref{fig:Population_Exp}, it is evident that \textbf{94\%} of the survey participants have engaged in software projects for a minimum of \textbf{10 years}, with their experience ranging from 10 to 41 years. This suggests that the results obtained in our study were derived from experts who possess extensive and valuable experience gained from a long and successful industrial career. Only \textbf{two} of the respondents reported having less than 10 years of experience, with 7 and 9 years respectively. 

We were fortunate to have received participation from a diverse group of architects in the bank, encompassing a broad range of roles. This enabled us to gain insights into software architecture from various levels of abstraction, as well as the experiences of different stakeholders. As illustrated in Figure \ref{fig:role-figure}, our participants spanned 7 different architectural roles, with the majority of them being Domain Architects (37\%) and IT Architects (26\%). The next most frequent roles were Solution Architects (11\%) and Enterprise Architects (8\%). 

\section{Results}
\label{sec:results}
The main results we inferred from the data\footnote{Raw Data of Survey: https://bit.ly/3EvC2H3} we gathered are presented in this section. It does so in accordance with the five blocks from \textbf{Q3 to Q7} and the survey structure specified in Section \ref{sec:design} of the report. 

\textbf{Table \ref{tab:ak_in_practice}} displays the responses to questions in the \textbf{Q3 block}, which focuses on the practical application of AK. We initiated the block by asking participants, "What does AK mean to you?" \cite{CAPILLA2016191}. Most participants provided definitions that complemented the formal definition we provided. For instance, one participant explained that \textit{"AK is a group of elements, such as problem statements, triggers to have a solution, reasons for creating a new solution, scope, requirements, assumptions, current state architecture, transition state, and target state. It involves defining and registering exceptions/deviations to deliver a solution building block."} However, a few participants shared unique perspectives. One stated that \textit{"AK is also about knowing what kind of architectural artifacts (e.g., future state architectures, current state architecture, guidelines, standards) exist in the organization and identifying any missing artifacts. But most importantly, it involves interpreting and using them correctly."}

We present the results for the \textbf{Q4 block}, which pertains to the practical representation of AK, in \textbf{Table \ref{tab:ic_criteria}}. \textbf{Table \ref{tab:AK_Communication}} displays the results of the \textbf{Q5 block}, which examines the implementation of AK communication. We began by asking participants about the stakeholders with whom they need to communicate AK knowledge in their present position. The stakeholders that participants engage with differ depending on their current role, in addition to their peers. For example, business architects communicate with business and IT management, business change teams, and IT delivery teams. Domain architects, on the other hand, engage with product owners, enterprise architects, principal architects, developers, business analysts, IT leads, Security CISO, Design Authority, and external vendors. IT architects communicate with Grid Owner, Product Owner, Business Analyst, Enterprise Architects, and Domain Architects.

The results for the \textbf{Q6 block}, which pertain to domain-specific versus domain-generic AK, are presented in \textbf{Table \ref{tab:domain_Specific_Generic}}. Finally, we summarize the results for the \textbf{Q7 block}, which concerns questions related to sustainability aspects within software architecture, in \textbf{Table \ref{tab:sustainability_aspect}}.

\begin{table}[!ht]
\caption{Architecture Knowledge (AK) representation in financial domain}
\label{tab:ak_in_practice}
\begin{tabular}{| m{0.03\textwidth} |  m{0.41\textwidth}|}
\hline
\multirow{2}{*}\textbf{Q2.2} & \textbf{Question:} What type of AK do you document and keep? Capilla et al.\cite{CAPILLA2016191}\\
\cline{2-2}
 & \textbf{Answer:} 26 out of 32 respondents, which represents 81\% of the participants, mentioned Solution Intent as the AK to keep and document. Meanwhile, Current State Architecture was mentioned by 10 participants (31\%), and Design Decisions were mentioned by 25\% of respondents. Future State Architecture was mentioned by 22\% of the participants. Five participants mentioned both Guidelines and Standards as the AK to capture and retain. Additionally, there were several other AK documents mentioned by the participants that are worth noting. These include High level Design, Information Architecture, Architectural Landscape, Problem Statements, and Architectural Review Comments.\\
\hline
\multirow{2}{*}\textbf{Q2.3} & \textbf{Question:} Do you capture AK consistently in all projects? Capilla et al.\cite{CAPILLA2016191} \\
\cline{2-2}
 & \textbf{Answer:} Of the 31 respondents to this question, 18 participants (58\%) answered "Yes" while the remaining 13 individuals (42\%) answered "No". Among the 13 who answered "No", some specified their reasons, including 31\% who said that every project is different, 31\% who stated that AK is not required, 15\% who believed that it can sometimes be an overkill, and others mentioned that it is labor-intensive and that they don't have enough time.\\
\hline
\multirow{2}{*}\textbf{Q2.4} & \textbf{Question:} In your experience, what are the AK elements that you miss?\\
\cline{2-2}
 & \textbf{Answer:} Of the 31 participants who answered this question, 14\% of them reported being satisfied with the existing AK elements. However, 26\% (8 individuals) identified AK elements that act as a means of communication and bridge between diverse stakeholders as the ones they miss the most in their work. On the other hand, 23\% pointed out that AK elements that give the context and status of the architecture were the ones they miss in their work.. Additionally, 13\% of participants mentioned missing AK elements related to Clarity, Detail, and Guidance, and 10\% mentioned missing elements related to Design Decisions. Finally, 6\% of participants missed AK elements related to Sustainability Aspects.\\
\hline
\multirow{2}{*}\textbf{Q2.5} & \textbf{Question:} In your experience, what are the AK elements that you find particularly useful?\\
\cline{2-2}
 & \textbf{Answer:} Out of 31 participants who responded to the question, 12 participants (39\%) found AK elements related to Standards, References, and Guidelines to be the most beneficial. Furthermore, 20\% of participants chose Architecture Model and Business Architecture as the most useful each. Thirteen percent of participants found Solution Intent\tablefootnote{A \textit{Solution Intent (SI)} describes the high level, implementation dependent solution architecture and provides guidance to teams by defining the solution direction.} to be the most beneficial, while 10\% each chose Design Decisions and context \& status of the architecture as the most useful AK elements.\\
\hline
\end{tabular}
\end{table}

\begin{table}[ht]
\caption{Architecture Knowledge (AK) representation in financial domain}
\label{tab:ic_criteria}
\begin{tabular}{| m{0.03\textwidth} |  m{0.41\textwidth}|}
\hline
\multirow{2}{*}\textbf{Q3.1} & \textbf{Question:} Do you know any standard notation or language to capture \textbf{AK}? Capilla et al.\cite{CAPILLA2016191}\\
\cline{2-2}
 & \textbf{Answer:} Of the 31 participants who responded to the question, 90\% (28 participants) indicated "Yes". Among these 28 participants, 82\% (23 individuals) specified that they use ArchiMate to capture AK, while 5 participants (18\%) specified using UML. Additionally, 2 participants each (totaling 7\%) mentioned using Sparx EA, Draw.io, PowerPoint, and BPMN as their preferred languages and notations for capturing \textbf{AK}.\\
\hline
\multirow{2}{*}\textbf{Q3.2} & \textbf{Question:}In your experience, what is the most useful architectural documentation? Capilla et al.\cite{CAPILLA2016191}\\
\cline{2-2}
 & \textbf{Answer:} Of the 31 participants who responded to this question, 9 individuals (29\%) identified ArchiMate as the most useful architecture documentation. Additionally, 5 participants each (totaling 16\% of the total) mentioned Current State Architecture and Diagrams as being useful. Three participants each (10\% of the total) identified Solution Intent, Views/Viewpoints, and PowerPoint as the most useful. Finally, 2 participants (7\% of the total) each mentioned Design Decision Documents and Visio as being useful.\\
\hline
\multirow{2}{*}\textbf{Q3.3} & \textbf{Question:} If you think about your last project, in what format was the knowledge stored?Capilla et al. \cite{CAPILLA2016191} \\
\cline{2-2}
 & \textbf{Answer:} 31 participants responded to the question on how they store AK, mentioning various formats and methods. ArchiMate, Word document, and PowerPoint were equally popular among 9 participants (representing 29\%) as the formats used to store AK in their last project. Solution Intent was mentioned by 8 participants (26\%), while Confluence was used by 6 participants (19\%).\\
\hline
\multirow{2}{*}\textbf{Q3.4} & \textbf{Question:} What tools or methods do you use to capture and represent AK?\\
\cline{2-2}
 & \textbf{Answer:} Of the 31 respondents, 17 individuals (55\%) use ArchiMate, 15 (48\%) use PowerPoint, 14 (45\%) use Sparx EA, and 11 (35\%) use Confluence to capture and represent AK. Additionally, 10 participants use Microsoft Word and 9 use Visio for this purpose.\\
\hline
\end{tabular}
\end{table}

\begin{table}[ht]
\caption{Architecture Knowledge(AK) communication in practice }
\label{tab:AK_Communication}
\begin{tabular}{| m{0.03\textwidth} |  m{0.41\textwidth}|}
\hline
\multirow{2}{*}\textbf{Q4.2} & \textbf{Question:}If you think of your last project and the AK you have stored, how did you communicate the knowledge to your stakeholders?\\
\cline{2-2}
 & \textbf{Answer:} In their last project, 26 participants responded to the question on how they communicated AK with different stakeholders. Among the respondents, 38\% (10 individuals) used Written Documents to communicate AK, while 35\% (9 individuals) mentioned using Meetings and PowerPoint equally. Additionally, Email was used by 27\% (7 individuals) of the participants, and Workshops were used by 3 individuals.\\
\hline
\multirow{2}{*}\textbf{Q4.3} & \textbf{Question:} In general, what tools or methods you use to share and communicate \textbf{AK}? \\
\cline{2-2}
 & \textbf{Answer:} Out of the 30 individuals who responded to this question, 47\% (14 participants) primarily use PowerPoint for communication, 37\% (11 participants) use email, 30\% (9 participants) use Confluence, and 20\% (6 participants) use both meetings and documents.\\
\hline
\end{tabular}
\end{table}

\begin{table}[ht]
\caption{domain-specific Vs generic-specific AK}
\label{tab:domain_Specific_Generic}
\begin{tabular}{| m{0.03\textwidth} |  m{0.41\textwidth}|}
\hline
\multirow{2}{*}\textbf{Q5.1} & \textbf{Question:} Do you know certain methods for \textbf{AK} which are exclusively valid or applied to your business domain?\\
\cline{2-2}
 & \textbf{Answer:} Out of 31 respondents to this question, only 5 (16\%) responded “Yes”, and specified BPMN, TOGAF, SAFe, Target State Design as exclusive AKs to the finance area.\\
\hline
\multirow{2}{*}\textbf{Q5.2} & \textbf{Question:} Can you think of any other AK methods that are general-purpose and that have not already been mentioned?\\
\cline{2-2}
 & \textbf{Answer:} In response to the question, 20 participants provided feedback. Of these, 11 participants (representing 55\% of the respondents) answered with "No." Among the 8 participants who provided an answer to the specific inquiry, they mentioned the following frameworks: TOGAF, Agile Design Thinking, BPMN, Service Oriented Architecture, and Standardized Architecture Decision Records (ARDs).\\
\hline
\multirow{2}{*}\textbf{Q5.3} & \textbf{Question:} In general, do you have to keep certain regulations in mind (e.g., GDPR, sustainability targets, etc.) while representing or communicating \textbf{AK} in your business domain? \\
\cline{2-2}
 & \textbf{Answer:} Out of 30 respondents to this question, 87\% (26 participants) disclosed that they consider certain regulations while representing and communicating AKs. \\
\hline
\end{tabular}
\end{table}

\begin{table}[ht]
\caption{Sustainability aspects of software architecture }
\label{tab:sustainability_aspect}
\begin{tabular}{| m{0.03\textwidth} |  m{0.41\textwidth}|}
\hline
\multirow{2}{*}\textbf{Q6.1} & \textbf{Question:} What is IT sustainability for you?\\
\cline{2-2}
 & \textbf{Answer:} Among the 23 individuals who responded to the question, 30\% (7 participants) demonstrated a comprehensive understanding by referring to at least two aspects of sustainability or software engineering principles. Specifically, 39\% (9 participants) associated IT sustainability with the environmental dimension, while 22\% (5 participants) focused on the technical dimensions. The remaining two participants understood IT sustainability in terms of its economic dimension and its ability to cope with changing environments across all domains.\\
\hline
\multirow{2}{*}\textbf{Q6.2} & \textbf{Question:}Are you aware of any IT related sustainability targets or measures in your organization/department?\\
\cline{2-2}
 & \textbf{Answer:} Out of 28 respondents to this question, 18 (representing 64\% of the total) reported not being aware of any sustainability targets. The remaining 10 participants (36\%) reported having knowledge of some targets, with the most commonly mentioned targets being cloud-related policies and targets related to energy consumption, both at 33\%.\\
\hline
\multirow{2}{*}\textbf{Q6.3} & \textbf{Question:} Do you consider sustainability aspects in your current role? \\
\cline{2-2}
 & \textbf{Answer:} Of the 28 participants who responded to the question, 75\% (21 participants) said "Yes". Out of these 21 participants, 48\% (10 participants) considered the technical aspect of sustainability, while 14\% (3 participants) considered the environmental aspect, and 9\% (2 participants) considered the economic aspect. Additionally, 14\% (3 participants) considered other aspects such as business and quality requirements, and adapting to changes.\\
\hline
\multirow{2}{*}\textbf{Q6.4} & \textbf{Question:} What tools or methods do you use to capture and represent AK?\\
\cline{2-2}
 & \textbf{Answer:} Of the 31 respondents, 17 individuals (55\%) use ArchiMate, 15 (48\%) use PowerPoint, 14 (45\%) use Sparx EA, and 11 (35\%) use Confluence to capture and represent AK. Additionally, 10 participants use Microsoft Word and 9 use Visio for this purpose. We also inquired about how and where they would incorporate sustainability into their daily work if it were necessary. Some suggested including it as a quality attribute in the Solution Intent or providing guidance on what to assess. Others suggested integrating it into the business architecture, domain architecture, intake phase, data center, and patterns and designs. \\
\hline
\end{tabular}
\end{table}

\begin{table*}[hbt!]
\caption{Description of Coding Categories of Qualitative data analysis}
\label{tab:description_of_coding_categories}
\begin{tabular}{|  m{0.18\textwidth}| m{0.28\textwidth}|m{0.36\textwidth}|m{0.04\textwidth}|m{0.04\textwidth}|}

\hline
\multicolumn{5}{|c|}{\textbf{Q.2.4 In your experience, what are the AK elements that you miss?}} \\ \hline
\textbf{Category} & \textbf{Description} & \textbf{Response Example} & \textbf{No} & \textbf{(\%)} \\ 
 \hline
 Communication \& Bridge & AK elements that serve as a bridge between various view/viewpoints intended for different stakeholders and enable effective communication to provide a comprehensive understanding of the architecture. & \textit{"As architecture is about communication, with different views, we tend to develop views that are understood by architects (which gives grip on the architecture), but less meaningful for (other) stakeholders, linking architecture views to (non architectural) views of different stakeholders is now lacking. We tend to speak our own language and not the language of different stakeholders."} & 8 & 26\% \\ 
 \hline
 Context \& Status & AK elements that provide actual status of IT components  and context of architectural changes & \textit{"Enterprise level Architectural Fitness Function, that are functions that can be added to the pipelines. So you can validate that what is implemented is in line with the architectural guidelines."} & 7 & 23\% \\ 
 \hline
 Clarity, Details \& Guidelines & AK elements related to meaning full principles, details, guidelines that apply to the entire IT landscape of the firm. & \textit{"The detail level of a solution design is always a challenge. Depending on what a team needs, you can decide to include more details. For teams that are mature less details are required. I think it would be good to have guidance on how detailed a SI should be."} & 4 & 13\% \\ 
 \hline
 Design Decisions & AK elements related to architectural design decisions. & \textit{"We have just started focusing on Design Decisions. We should focus more on this. DD captures clear rationale for a specific choice and the other alternatives assessed including pros and cons. This helps a person also understand how a system has evolved and why it has evolved this way. It may point to improvements that are possible today and not when the DD was made."} & 3 & 10\% \\ 
 \hline
 Sustainability Aspects & AK elements related to sustainability aspects, targets, and goals. & \textit{"sustainability aspect of a solution"} & 2 & 6\% \\ 
 \hline
 \multicolumn{5}{|c|}{\textbf{Q.6.1 What is IT sustainability for you?}} \\ \hline
 Environmental & Response  that focuses on reducing an organization’s impact on the environment such as lowering carbon emission, improving water and air quality. & \textit{"IT sustainability is taking the effects into account that an IT implementation has on energy consumption and environment. Considering how to reduce this footprint for generations to come.  
"} & 9 & 39\% \\ 
 \hline
  Comprehensive & A description that consists of two or more dimensions of sustainability or principles
of software engineering. & \textit{"Three parts: 1) archtecture in such a way that the solution requires as less resources (and thus energy) as possible and 2)define and develop in such a way that solution can easily be replaced or enhance when requirements change and 3) set-up a sustainable business model."} & 7 & 30\% \\ 
 \hline
 Technical & Responses that focused on how software and hardware can be designed and developed to reduce their environmental impact and improve their efficiency& \textit{"Reduce complexity; reuse of solutions; run systems on-demand; agile architectures"} & 5 & 22\% \\ 
 \hline
\end{tabular}
\end{table*}

\section{Discussion}
\label{sec:discussion}

We conducted a survey with software architects in one of the leading banks in the Netherlands and identified two main findings, along with a list of detailed results from analyzing each cluster of questions designed to address our research question as discussed in section 3. Overall, the survey yielded valuable insights into the representation and communication of software architecture knowledge in practice. The findings are: 

\begin{tcolorbox}[enhanced,attach boxed title to top center={yshift=-3mm,yshifttext=-1mm},
  colback=blue!5!white,colframe=blue!75!black,colbacktitle=red!80!black,
  title=\textit{Finding 1},fonttitle=\bfseries,
  boxed title style={size=small,colframe=red!50!black} ]
 A new architectural element that links the various architectural features and viewpoints created for various stakeholders appears to be necessary.
\end{tcolorbox}

During the survey, we asked participants about the architectural elements they felt were missing in their work. This question was found to be revealing as it elicited many interesting responses. We categorized and analyzed these responses, which can be seen in Table \ref{tab:description_of_coding_categories}, to facilitate quantitative data analysis. Most participants identified the need for architectural elements that can facilitate communication and bridge different viewpoints and features for various stakeholders. For example, one participant stated, \textit{"As architecture is about communication, with different views, we tend to develop views that are understood by architects (which gives grip on the architecture), but less meaningful for (other) stakeholders. Linking architecture views to (non-architectural) views of different stakeholders is now lacking. We tend to speak our own language and not the language of different stakeholders."} This view expressed by a participant is not isolated, as shown in Figure 5, where 8 out of 31 respondents (26\%) shared similar views. Another participant expressed the desire for "more linkage to use non-architecture viewpoints" to better represent and communicate AK.

\begin{figure}[htbp]
\center
\includegraphics[width=1.0\columnwidth]{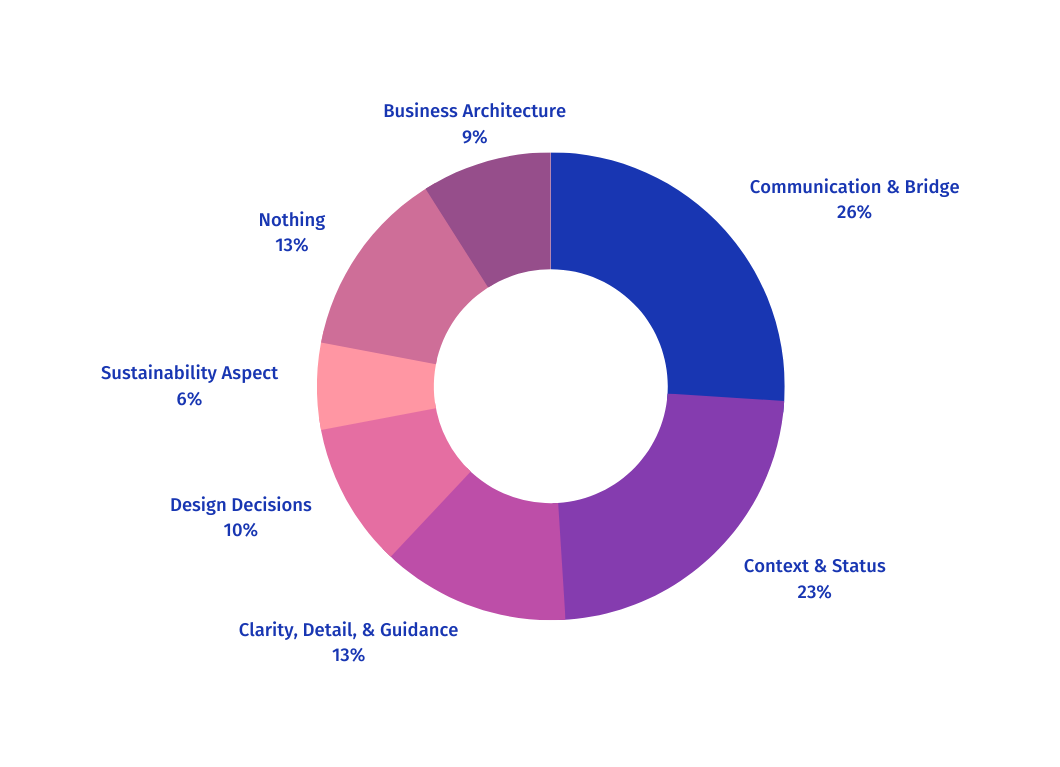}
\caption{Architectural Elements missed by Architects}
\label{fig:architectural-elements}
\end{figure}

  \begin{tcolorbox}[enhanced,attach boxed title to top center={yshift=-3mm,yshifttext=-1mm},
  colback=blue!5!white,colframe=blue!75!black,colbacktitle=red!80!black,
  title=\textit{Finding 2},fonttitle=\bfseries,
  boxed title style={size=small,colframe=red!50!black} ]
 Clear guidance, references, and goals are necessary to motivate architects to practice Sustainable Software Engineering.
\end{tcolorbox}

\begin{figure}[htbp]
\center
\includegraphics[width=1.0\columnwidth]{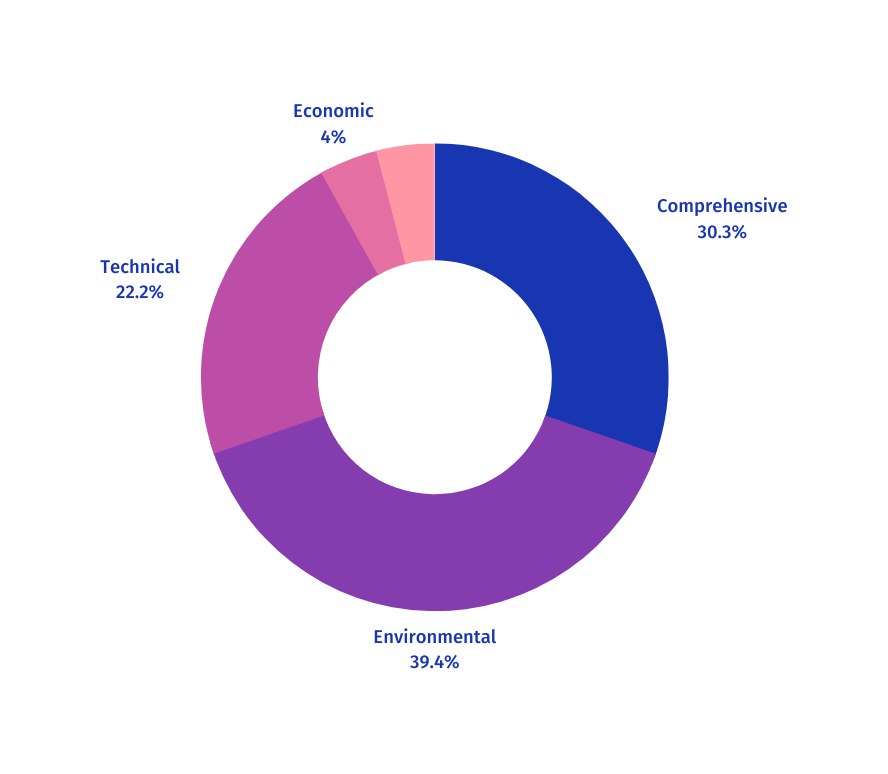}
\caption{Architects understanding of Sustainable IT }
\label{fig:sustainability-awareness}
\end{figure}

Our study has yielded significant insights into the sustainability aspects of software architecture, with a particular focus on representation and communication. To assess the sustainability awareness of architects, we asked them to define sustainable IT. Our analysis revealed that one-third of the participants who responded demonstrated an impressive understanding by mentioning two or more aspects of sustainability or sustainable software engineering principles, as depicted in Figure \ref{fig:sustainability-awareness}. Moreover, other participants referred to various sustainability dimensions, indicating a high level of awareness on the subject.

Subsequently, we inquired whether participants were aware of their organization or department's sustainability targets, and the majority responded negatively, highlighting a lack of awareness in this regard. Nevertheless, when we asked whether they integrate sustainability aspects into their work, 75\% of participants responded affirmatively, which appears to contradict their lack of awareness of sustainability targets. However, this discrepancy may be attributed to their advanced level of understanding of the concept.

Upon further investigation, most participants reported that they were not incorporating sustainability aspects into their daily work due to a dearth of clear guidelines, references, criteria, and goals. Additionally, they expressed a need for more support in bridging the knowledge gap on how to implement sustainability aspects in their work. Overall, our study underscores the importance of incorporating clear guidelines, references, criteria, and goals on sustainability aspects by the organization to leverage the motivation and high level of sustainability awareness of architects.

Two main research questions and three sub questions served as the framework for the survey that is the subject of this report. The following provides a summary of the findings.

\begin{tcolorbox}[colback=blue!5!white,colframe=blue!75!black,colbacktitle=red!80!black]
  RQ1\textit{ How is software architecture knowledge represented and communicated in practice?}
\end{tcolorbox}
During our research, we discovered that architecture knowledge (AK) is communicated and represented through various documentation and artifacts such as Solution Intent, Current State Architecture, Design Decisions, and Guidelines. However, not all projects consistently capture AK, and some participants mentioned missing AK elements related to communication, context and status,  design decisions, and sustainability aspects. On the other hand, participants found AK elements related to standards, references, guidelines, architecture models, and business architecture to be particularly useful.

\begin{tcolorbox}[colback=blue!5!white,colframe=blue!75!black,colbacktitle=red!80!black]
  RQ1.1\textit{ How is software architecture knowledge represented in the financial domain?}
\end{tcolorbox}
Based on the responses, it can be concluded that standard notations or languages, such as ArchiMate and UML, are commonly used to capture and represent software architecture knowledge in the financial domain. ArchiMate was reported as the most commonly used notation. Various tools, including PowerPoint, Sparx EA, Confluence, Word documents, and Visio, were identified as useful for capturing and representing architecture knowledge. The most useful architectural documentation included ArchiMate, current state architecture and diagrams, solution intent, views/viewpoints, and PowerPoint.

\begin{tcolorbox}[colback=blue!5!white,colframe=blue!75!black,colbacktitle=red!80!black]
  RQ1.2\textit{ How is software architecture knowledge communicated in the financial domain? }
\end{tcolorbox}
The findings suggest that software architecture knowledge in the financial domain is communicated through various methods and tools, including written documents, meetings, presentations, email, and workshops. The choice of communication method may depend on the stakeholder, the level of involvement, and the complexity of the information. Overall, PowerPoint is the most commonly used tool for sharing and communicating AK, followed by email and Confluence. Meetings and documents are also frequently used, with some participants reporting the use of workshops. However, it is important to note that the specific methods and tools for communicating software architecture knowledge may differ depending on the industry or domain.

\begin{tcolorbox}[colback=blue!5!white,colframe=blue!75!black,colbacktitle=red!80!black]
  RQ1.3\textit{ What architecture knowledge methods are domain-specific or domain-generic?}
\end{tcolorbox}
Architecture knowledge (AK) methods are employed in practice through a combination of domain-specific and domain-agnostic approaches. While most respondents did not identify any AK elements specific to the finance domain, a few mentioned techniques such as BPMN, TOGAF, SAFe, and Target State Design as being exclusive to finance. However, it is worth noting that TOGAF and BPMN are not solely utilized in finance. Only a few general-purpose AK frameworks such as Agile Design Thinking, Service Oriented Architecture, and Standardized Architecture Decision Records (ARDs) were mentioned. This suggests that there may be a lack of awareness of domain-specific AKs or that many methods are considered applicable across different domains. Notably, 84\% of respondents did not mention any domain-specific AKs, highlighting the need for further exploration of the AK methods unique to specific domains.

\begin{tcolorbox}[colback=blue!5!white,colframe=blue!75!black,colbacktitle=red!80!black]
  RQ2\textit{ How can sustainability aspects be represented and communicated in software architecture?}
\end{tcolorbox}
To incorporate sustainability aspects into software architecture, clear guidelines, references, and goals are required to capture these aspects in daily work. Participants suggest integrating sustainability into the business and domain architecture, intake phase, data center, and patterns and designs. Quality attributes in the Solution Intent can also be used to represent sustainability aspects. However, a comprehensive understanding of the various dimensions and principles of sustainable software engineering is necessary for effective representation and communication. Providing clear guidance, references, and goals can motivate architects to practice sustainable software engineering and integrate sustainability into their daily work. 

\section{Threats to Validity}
\label{sec:validity_threat}

While we made every effort to conduct a rigorous study, there are several limitations that must be acknowledged. The \textbf{\textit{external validity}} of our study may be limited by the fact that we only targeted participants from a single organization, despite the fact that we were able to attract a decent number of participants with significant experience in software architecture. Although the organization we targeted was large and likely representative of the industry as a whole, the lack of diversity in our population may limit the generalizability of our findings. As a result, it may be necessary to replicate our study with a more diverse sample in order to confirm the external validity of our results.

The \textbf{\textit{internal validity}} of a study can be threatened by various factors that impact the accuracy and reliability of the study's results. One potential threat to internal validity in a survey is the use of non-mandatory questions. In our study, we designed most of the questions to be non-mandatory to avoid obliging participants to answer questions they may not be qualified to answer or may have distaste towards. However, this design choice can impact the overall quality of responses received, as participants may choose not to answer certain questions, resulting in missing data and potentially biased results. To address this internal validity threat, we took a careful approach to analysing the survey responses. Rather than using the total recorded response for each question, we only considered the total number of respondents who answered each specific question. By doing this, we were able to account for missing data and ensure that the responses analysed for each question were only from participants who chose to answer that particular question. This approach allowed us to mitigate the potential impact of non-mandatory questions on the study's internal validity and ensure that our results were as accurate and reliable as possible.

\textbf{\textit{Construct validity}} is a critical aspect of any research study that seeks to examine and measure theoretical concepts or constructs. In our study, we aimed to explore the perception of software architecture and architectural knowledge related to sustainability aspects, and we focused on software architects with a lot of experience to gather insights. While software architects may be the ideal candidates to respond to questions related to software architecture, it can be challenging to determine the best approach for measuring and analyzing sustainability aspects in software architecture due to the lack of an established view on the combination of these two areas. As researchers, we made every effort to define the theoretical concepts and constructs we wished to study and determine how to measure them in a valid and reliable way. However, the lack of consensus on the combination of sustainability and software architecture posed a significant challenge in this regard. Therefore, we opted to investigate how architects perceive sustainability concepts and where they may apply sustainability to address sustainability in software architecture. This approach allowed us to explore the perceptions and perspectives of experienced software architects, even in the absence of a well-established theoretical framework for the combination of sustainability and software architecture. However, this construct validity threat must be considered when interpreting our study's findings, and further research is needed to establish a more robust theoretical foundation for the study of sustainability in software architecture.
\section{Conclusion and Future Work}
\label{sec:conclusion}

This paper presents the findings of a survey we conducted on the representation and communication of architectural knowledge (AK) in practice. Our study targeted software architects working for a leading bank in the Netherlands with extensive industry experience in various architectural roles. We identified two main findings through our analysis of the survey results: the need for a new architectural element that links different features and viewpoints created for various stakeholders, and the need for clear guidance, references, and goals to motivate architects to practice sustainable software engineering. These findings offer valuable insights for future research in the field. We recommend further investigation into the development of this new architectural element and how it can be integrated into existing practices. Additionally, we suggest exploring ways to promote sustainable software engineering practices among architects through the establishment of clear guidance and goals. Our study highlights the importance of effective AK representation and communication in software industry and the potential benefits of incorporating sustainable practices into architectural decision-making.

\bibliographystyle{ACM-Reference-Format}
\bibliography{biblio}

\end{document}